\documentclass[twocolumn]{aastex631}
\usepackage{lipsum}
\usepackage{amsmath}
\usepackage{mathtools}
\usepackage{soul}
\usepackage[nottoc]{tocbibind}
\usepackage{hyperref}
\shorttitle{SBO}
\begin{document}
\title{Ultraviolet radiation and neutrinos: two messengers from CCSNe in the CSM scenario}
 
\correspondingauthor{silviagagliardini@hotmail.com}

\author[0000-0002-1832-2126]{Silvia Gagliardini}
\affil{Physics Department,Ariel University, Ariel, Israel}
\affil{Istituto Nazionale di Fisica Nucleare, Sezione di Roma, P. le Aldo Moro 2, I-00185 Rome, Italy}
\affil{Dipartimento di Fisica dell'Universit\`a La Sapienza, P. le Aldo Moro 2, I-00185 Rome, Italy}
\author[0000-0003-4366-8265]{Simone Dall'Osso}
\affil{INAF - Istituto di Radioastronomia, via Gobetti 101, Bologna, 40129, Italy}
%\affil{Dipartimento di Fisica e Astronomia -- Universit\`{a} di Bologna - Bologna, Italy}
\author[0000-0002-7349-1109]{Dafne Guetta}
\affil{Physics Department,Ariel University, Ariel, Israel}
\author[0000-0000-0000-0000]{Angela Zegarelli}
\affil{Ruhr University Bochum, Faculty of Physics and Astronomy, Astronomical Institute (AIRUB), Universitätsstraße 150, 44801 Bochum, Germany}
\author[0000-0000-0000-0000]{Silvia Celli}
\affil{Istituto Nazionale di Fisica Nucleare, Sezione di Roma, P. le Aldo Moro 2, I-00185 Rome, Italy}
\affil{Dipartimento di Fisica dell'Universit\`a La Sapienza, P. le Aldo Moro 2, I-00185 Rome, Italy}
\author[0000-0000-0000-0000]{Antonio Capone}
\affil{Istituto Nazionale di Fisica Nucleare, Sezione di Roma, P. le Aldo Moro 2, I-00185 Rome, Italy}
\author[0000-0000-0000-0000]{Irene Di Palma}
\affil{Istituto Nazionale di Fisica Nucleare, Sezione di Roma, P. le Aldo Moro 2, I-00185 Rome, Italy}
\affil{Dipartimento di Fisica dell'Universit\`a La Sapienza, P. le Aldo Moro 2, I-00185 Rome, Italy}

\submitjournal{ApJ}

\begin{abstract}

Massive stars (\(>8\,M_{\odot}\)) often undergo intense mass loss through winds or eruptive events in the final stages of their evolution, leading to the formation of a dense circumstellar medium (CSM). This material, expelled months to years before core collapse, shapes the pre-explosion environment and influences the early supernova (SN) emission.~In particular, the interaction of the SN ejecta with the dense CSM can power an extended emission  
into the UV/optical bands, as seen in a growing fraction of type II SN.~Recent events such as SN~2023ixf and SN~2024ggi confirm the relevance of dense environments and highlight the value of UV observations.~Moreover, Fast Blue Optical Transients (FBOTs) may represent extreme cases of this interaction, possibly linked to more compact/massive CSM.  In this work, we model %SBOs 
the SN-CSM shock interaction in order to (i) estimate the maximum detection horizons and expected rates for future UV missions like ULTRASAT, and (ii) to estimate the intensity and expected rate of potential neutrino signals detectable by IceCube and KM3NeT. We then
discuss the prospects for multi-messenger observations of such events in the near future.

\end{abstract}

\section{Introduction}
During the terminal phases of stellar evolution, massive stars ($>8 M_{\odot}$) can undergo substantial mass loss via intense stellar winds or eruptive events.~These episodes, which may occur months to years prior to core-collapse, lead to the formation of a dense circumstellar medium (CSM) that significantly alters the immediate pre-supernova 
environment.~Type II Supernovae (SNe~II) are thought to arise from core-collapse explosions of red supergiants (RSGs), a connection confirmed by the identification, via pre-explosion imaging, of progenitor stars for dozens SNe II (\citealt{Smartt_2015}), including SN 2017eaw 
\citep{Kilpatrick2018,Van_Dyk_2019,Rui2019}, and SN 2023ixf 
\citep{Kilpatrick_2023,Jencson_2023,Soraisam_2023}.~Some progenitors were found to be surrounded by thick CSM or dust shells, resulting in suppressed flux at optical wavelengths and enhanced emission in the infrared (IR), as observed in SN 2012aw, SN 2017eaw, and SN 2023ixf \citep{Xiang_2024, li_2024}.

The earliest emission from core-collapse supernovae (CCSNe) %supernova (SN) explosion 
is a so-called `shock breakout' (SBO).~As the radiation-mediated shock (RMS) (\citealt{1976ApJS...32..233W}) responsible for ejecting the stellar envelope moves outward, the optical depth of the overlying plasma decreases.~
 The shock becomes optically thin as it approaches the surface, and releases a
%While approaching  the stellar surface 
%$\tau$ drops to  $\sim c/v$, $v$ being the shock velocity,
%
%the shock becomes optically thin releasing a %brief and 
bright X-ray/UV flash lasting a few hours \citep{Lasher_Chan_1975,Klein_Chevalier_1978}.%~This typically happens as the shock approaches the stellar surface \citep{Lasher_Chan_1975,Klein_Chevalier_1978}.
~The SBO  is followed by the standard supernova UV/optical emission from the cooling, expanding shocked envelope over a timescale of several days.

The initial burst of radiation from the SBO at the stellar surface can rapidly ionize the surrounding CSM, leading to distinctive spectral signatures observable in the SN light curve.~Type IIn supernovae, in particular, exhibit narrow emission lines produced by the slowly expanding, photoionized CSM enveloping the progenitor star (e.g., \citealt{Dessart_2015, Ercolino_2024}).~Similar signatures of a dense CSM have recently been observed in SN2024ggi like.~Early-time ("flash") spectroscopy, taken within 0.8 days of discovery, reveals emission lines from HI, HeI, CIII, and NIII, featuring a narrow core with broad, symmetric wings, characteristics similar to those of Type IIn supernovae \citep{Jacobson-Galán_2024}. 

Independent evidence for the prevalent presence of optically-thick CSM shells around spectroscopically regular SNe II progenitors is obtained from early, 1 day time scale observations of optical-UV SN light curves \citep{Morag_2023,Irani_2024}.%conducted the first systematic investigation of early ($\sim$ 1 day) optical-UV light curves across an extensive sample of Type II supernovae, expanding upon earlier analyses focused on individual objects (e.g. \citealt{Morag_2023}). 
~Studying a sample of SNe II, it was found that, while ~50\% 
of the early lightcurves are consistent with the emission from the expanding shocked stellar envelope, so-called “shock cooling” emission (see \citealt{waxman17}; \citealt{levinson_2020} for reviews), the light curves of the other 50\% are inconsistent with shock cooling and indicate the presence of an optically thick CSM shell \citep{Waxman_2024}.

%A new class of optical transients has gained attention in recent years:~Fast Blue Optical Transients (FBOTs).~They are characterized by rapid rise and decline timescales, typically just a few days, and high luminosities, on the order of $10^{43}$ erg/s, with emission peaking in the blue and ultraviolet bands \citep{Drout_2014, Guarini_2022}.~
The UV light curves of CCSNe typically decline on a timescale of days as the photosphere cools \citep{Brown_2009,Pritchard_2014}.~The collision of the SN ejecta with a dense CSM shell, though, generates a new shock which will power emission across a broad range in the EM spectrum.~In particular, it will significantly prolong the UV-bright phase beyond photospheric cooling %phase 
\citep{Fransson_1984,Chevalier_1994}.~Detection of such a prolonged emission will therefore indicate the additional energy provided by the SN-CSM interaction.~Moreover, the recently discovered class of Fast Blue Optical Transients (FBOTs) has gained much attention in recent years, sparking interest for a possible association with core-collapse explosions within a dense CSM environment.~FBOTs are characterized by rapid rise and decline timescales, typically only a few days, and high luminosities typically $\sim$ $10^{43}$ erg/s, with emission peaking in the blue and UV bands \citep{Drout_2014, Guarini_2022}.~It 
%The interaction between the supernova ejecta and {\color{orange} a dense CSM shell} can lead to higher temperatures and a rapid rise in the early light curves.~
%Interestingly, the erendipitous capture of pre- and post-explosion data from SN 2023ixf revealed that the shock wave passed through a dust shell, leading to a quick change of shock radiation from red to blue light \cite{li_2024}. 
%When the shockwave from the supernova interacts with the surrounding material, it generates intense UV radiation. 
is therefore crucial to be able to detect the UV emission in order to understand the shock dynamics and the properties of the CSM, as well as to shed light on newly-discovered classes of astrophysical transients.

The UV spectral range remains insufficiently covered by currently operational instruments.~The Ultraviolet Transient Astronomy Satellite (ULTRASAT; \citealt{ultrasat}), scheduled for 2027, will explore the sky in the UV band ($\lambda_{\rm min}$ = 230 nm; $\lambda_{\rm max}$= 290 nm). One of the main science cases for ULTRASAT is the early~(hrs) detection of core-collapse supernovae~(CCSN) and the high cadence (minutes) monitoring of their UV light curves. ULTRASAT is designed to be sensitive to the near-UV (NUV) regime and will be equipped with a UV telescope offering an unprecedented field of view (FoV) of 204 deg$^2$.

Collisionless SN shocks driven into the winds around massive stars and in the interstellar medium (ISM) have long been suggested as sources of high energy Cosmic Rays (CRs)
\citep{BLANDFORD19871}, and therefore are also considered as candidate sources of  high energy neutrinos \citep{Murase_2011, Petropoulou_2017, Sarmah_2022}.~The shock interaction of SN ejecta with a thick CSM is expected to lead to orders of magnitude enhanced neutrino luminosity, relative to the interaction with regular RSG winds (\citealt{Waxman_2024}).~In particular, the large neutrino luminosity would be produced during and shortly after the  break-out of the shock  through the dense CSM shell \citep{Li_2019, rw11}. 

Observations of such high-energy neutrinos can provide insights into the non-thermal processes occurring during the shock evolution.~Several authors (\citealt{margalit_2022}; \citealt{Waxman_2024}) have shown that a particularly promising observing scenario involves the detection of early UV emission, which is the first electromagnetic (EM) signal to emerge after the explosion. 
We therefore suggest that the UV flash will act as an effective trigger
for multi-messenger (MM), in particular neutrinos, and multi-wavelength follow-up observations. This would represent a fundamental breakthrough in MM astronomy and in the study of SNe.

In this work we focus on the earliest signals from the SN-CSM interaction 
and the associated neutrino emission. Our study forms part of a broader investigation dedicated to exploring the potential of ULTRASAT to observe CCSNe. In this paper, we expand on a previous study \citep{art:zegarelli} concerning UV emission from choked jets in CCSNe (SN IIL, IIP), where the interaction was confined to the extended hydrogen envelope of the progenitor star. We model the shock interaction of the SN ejecta with a dense CSM shell \citep{margalit_2022} to estimate the maximum %We model SBOs from CCSNe interacting with a CSM \citep{margalit_2022} to estimate the maximum
detection horizon for such events with ULTRASAT, and their corresponding detection rate. Moreover, we explore the prospects for MM detection of CCSNe interacting with a CSM, focusing on their UV and neutrino signatures as observable by ULTRASAT, IceCube, %a cubic-kilometer neutrino detector located at the South Pole, 
and KM3NeT, a next-generation observatory currently in operation while under construction in the Mediterranean Sea. These facilities are sensitive to high-energy neutrinos, making them particularly well suited to probing shock-powered transients embedded in dense CSM environments.
Recent studies have investigated neutrino production in CCSNe with dense CSM, highlighting their potential as sources of both low- and high-energy neutrinos \citep{Murase_2011, murase_2013, murase_2023, Murase_2024, Guetta_2023}.
%These include theoretical predictions and observational constraints (e.g., \citealt{Murase_2011, murase_2013, murase_2023, Murase_2024}), as well as dedicated case studies such as SN~2023ixf \citep{Guetta_2023}.
~These studies provide a solid foundation for assessing the MM detectability of these events with current and upcoming observatories.

The paper is organized as follows: In Section \ref{sec:SBO}, we describe the model used to calculate the expected luminosity of CSM-interacting type-II SNe; in Section \ref{sec:ULTRASAT} we discuss the ULTRASAT observational horizon. For comparison, we also evaluate the ZTF horizon in the context of our model. In Section \ref{sec:rate}, we present the expected rate of CCSNe. Section \ref{sec:neutrino} covers the expected neutrino emission from these events. Finally, Section \ref{sec:discussion} offers a discussion and summary of the findings.

\section{CSM-driven SBO light curves}

\label{sec:SBO}

Analytic solutions for the bolometric light curve produced by a dense, shocked CSM shell were derived by \cite{margalit_2022}. We adopted these solutions to calculate the expected luminosity of CSM-interacting type-II SNe in the ULTRASAT band, and to derive the maximum distance at which they can be detected.

The model assumes a spherical ejecta expanding with velocity $v_0$, which runs into a dense shell of CSM located at radius $\sim R_{\rm 0}$, with initial width $\Delta R_{\rm csm}$ and initial optical depth $\tau_0 > c/v_0$. 

As the spherical ejecta sweep the CSM shell, a shock is formed and part of the ejecta's kinetic energy is dissipated into post-shock CSM thermal energy. Because $\tau_0 > c/v_0$, radiation is initially trapped within the flow, and the shock will be mediated by radiation pressure. 

The CSM shell is thus accelerated, heated, and swept by the shock into a thin shell of width $\Delta R_0 = \Delta R_{\rm csm}/7\footnote{For a radiation-dominated shock with adiabatic index $\gamma = \frac{4}{3}$, the Rankine-Hugoniot conditions predict a compression ratio of 7, causing the post-shock shell thickness to be one-seventh of the pre-shock CSM width.} \ll R_0$.
The time-dependent bolometric luminosity, \( L(t) \), emerging from the outer edge of the CSM, depends on the explosion energy, \( E_0 \), which sets the overall normalization, as well as on three characteristic timescales: \( t_{\rm dyn} \), \( t_0 \), and \( t_a \).
The dynamical timescale, \( t_{\rm dyn} \equiv %\displaystyle \frac{R_0}{v} 
R_0/v \), 
represents the time required for the CSM to double its initial radius and $v$ is the expansion velocity of the outer edge of the shell (the maximum velocity throughout the shell).\\
The initial expansion timescale, \( t_0 \equiv %\frac{\Delta R_0}{v} 
\Delta R_0/v \), characterizes the time it takes for the CSM to expand across its initial width. The Arnett time, \( t_a \), is a characteristic diffusion timescale over which the shock-deposited energy is radiated away. It determines the overall duration of the transient and approximately corresponds to the time of peak luminosity. Finally, the light-curve shape is influenced by the velocity ratio $\beta \equiv v_0/v$ between the shell's inner and outer edge, respectively. 
The resulting analytical expression for the bolometric luminosity is:

\begin{multline}
    L(t)= \frac{7^2}{6^3} \beta \frac{E_0}{t_{\rm dyn}^2} \frac{(t_{\rm dyn}+t_0 +t)^2}{t_0 +(1-\beta)t} \times \\ \left[ \frac{(t_{\rm dyn}+t_0+t)^3 -(t_{\rm dyn}+\beta t)^3}{(t_{\rm dyn}+t_0)^3-t_{\rm dyn}^3}\right]^{-1/3} \times \\
    \sum_{n=1}^{\infty} \frac{4}{(2n-1)\pi} \sin\left[ \frac{2n-1}{2} \pi \frac{6^3}{7^2 \beta}\left(\displaystyle \frac{t_{\rm dyn}}{t_a}\right)^2\right] \times \\
    \left[1+(1-\beta) \frac{t}{t_0}\right]^{-\left(\displaystyle \frac{2n-1}{2}\pi\right)^2 \left(\displaystyle \frac{t_{\rm dyn}}{t_a}\right)^2 \displaystyle \frac{(1-\beta -\beta t_0/t_{\rm dyn})^2}{1-\beta}}
    \times \\e^{{ -\left(\displaystyle \frac{2n-1}{2}\pi\right)^2 \displaystyle \frac{t[(1-\beta^3)t+(2-4 \beta(\beta+1))t_0 + 6(1-\beta^2)t_{\rm dyn}]}{6(1-\beta)^2 t_a^2}}} \, ,
    \label{eq:lumi}
\end{multline} where the total initial internal energy is \[E_0 = \frac{M v_0^2}{2},\] and all the parameters in Eq.~\ref{eq:lumi} are directly related to the physical properties of the CSM shell, namely its mass \( M \), initial radius \( R_0 \), and the expansion velocity extrema, $v_0$ and $v$.~In this work we
adopt fiducial values for $\Delta R_0/R_0 =1/7$, $\beta = 1/2$ and $v_0 = 10^9~{\rm cm/s}$, and study the UV emission from the SBO as a function of the CSM shell mass ($M$) and radius ($R$), which are allowed to vary freely within physically plausible ranges, respectively
$R_{\rm 0}=(10^{13} \div 10^{15}){\rm cm}$ and $M=(10^{-3} \div 0.4)M_{\odot}$.~

A physical requirement for the radiation to remain initially trapped within the shell is that the optical depth satisfies \( \tau_0 > c / v_{\rm sh} \), where \( v_{\rm sh} \) is the velocity of the RMS propagating through the material. This condition imposes a constraint on the allowed parameter space, which can be expressed as %$\frac{
$t_{\rm dyn}/t_a < 0.5\, \beta^{1/2}$, as derived by \citet{margalit_2022}.

In the original work by \cite{arnett1980, arnett1982}, higher-order ($n > 1$;~Eq.~\ref{eq:lumi}) spatial modes were neglected, but the importance of including higher-order terms was later emphasized by \cite{Pinto_2000}, even in the %emphasized the importance of including these higher-order terms and  
simple case of a uniform energy distribution within the shock.~The model proposed by \cite{margalit_2022}, which assumes a fully uniform initial energy distribution, results in a divergent flux at t=0.~To address this unphysical behavior, \cite{margalit_2022} adopted a finite series expansion to approximate a uniform energy distribution, and demonstrated that eq. \ref{eq:lumi} converges smoothly in this case, if terms up to n=125 are considered: only modes of even higher order are effectively negligible. In our computations, we set the number of terms in the series to n=125, which ensures adequate convergence of the solution. 
While the truncated solution in Eq.~\ref{eq:lumi} appears to diverge as \( t \to 0 \) when considering only low-order terms (e.g., the \( n = 1 \) mode), \citet{margalit_2022} demonstrated that the full solution, which assumes a spatially uniform initial energy distribution, does in fact converge smoothly in this limit, provided that terms up to \( n = 125 \) are included. Contributions from higher-order modes become negligible beyond this point.

\begin{figure}
    \centering
    \includegraphics[width=1.1 \linewidth]{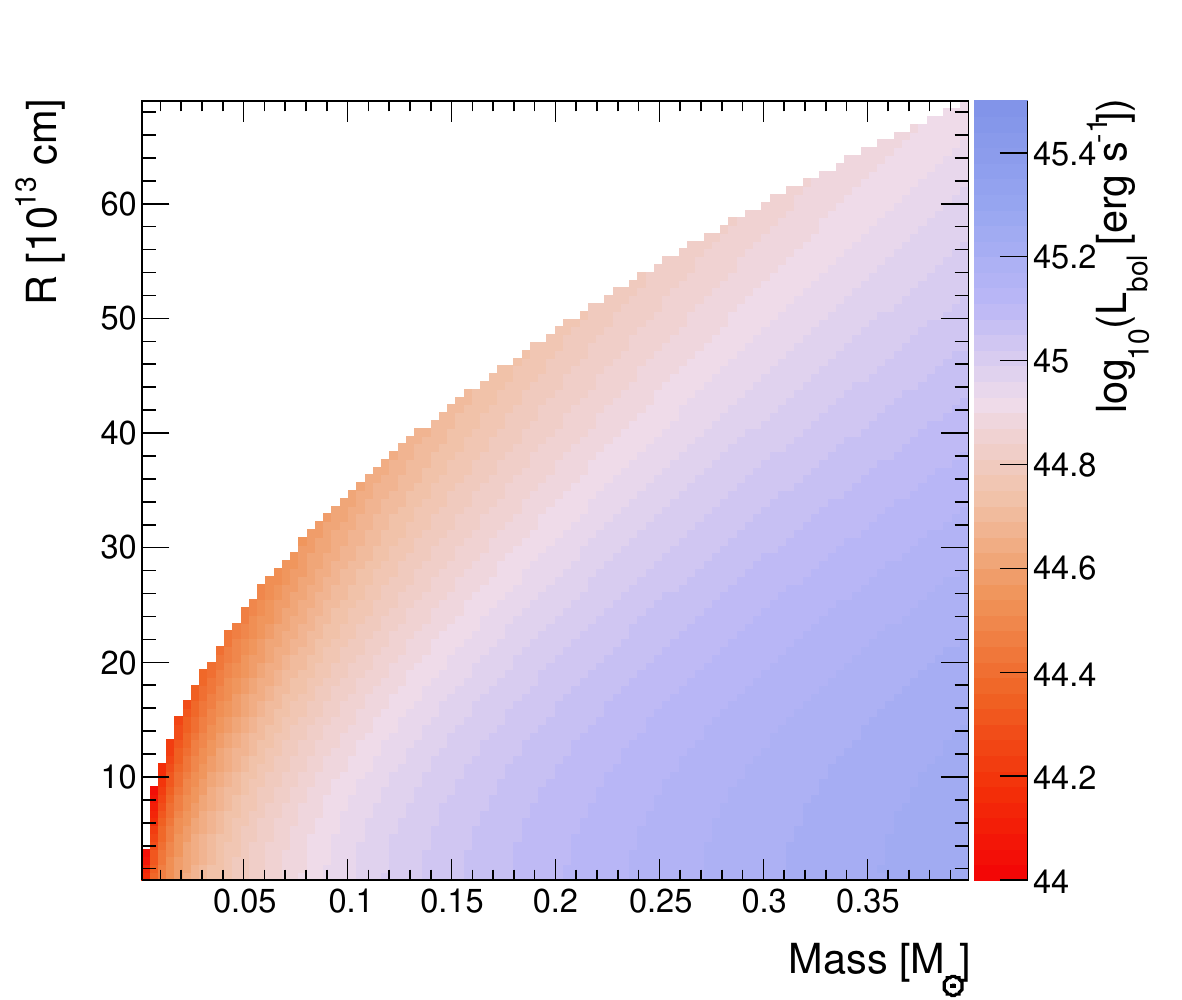}
    \caption{Bolometric luminosity of SNinteracting with circumstellar shells, computed for CSM masses ranging from $10^{-3}\,M_\odot$ to $0.4\,M_\odot$ and inner shell radii between $10^{13}$~cm and $10^{15}$~cm.}

    \label{lumi_bolo}
\end{figure}

~Accordingly, in our computations we always expanded Eq.~\ref{eq:lumi} up to the n=125 term. 
In Figure~\ref{lumi_bolo} we show the result of our model calculations for the expected bolometric luminosity as a function of the CSM mass (M) and initial radius (R).

The figure explores how variations in these parameters influence the luminosity output, under the assumption of a spatially uniform initial energy distribution. As the CSM mass increases or the shell extends over a larger radius, the interaction between the ejecta and the CSM becomes more efficient, leading to a brighter and more prolonged radiative signal.

\section{ULTRASAT DETECTION HORIZON}

\label{sec:ULTRASAT}

In order to estimate the detectability with ULTRASAT of SBO radiation from CSM-interacting SNe, we calculated  which fraction of the bolometric luminosity is emitted in the observing band (230-290 nm), given the expected blackbody emission spectrum.~Moreover, in order to calibrate our model and compare performances, we repeated the calculation also for the ZTF. 

The SBO luminosity in the ULTRASAT band, $L_{\rm UV}(t)$, can be expressed as a fraction $f_{\rm UV}(t)$ of  $L_{\rm SBO}(t)$,~with

\begin{equation}
f_{\rm UV}(t,z)= \displaystyle \frac{\int_{\lambda_{\rm min}/(1+z)}^{\lambda_{\rm max}/(1+z)}B_{\lambda}\left[T_{\rm bb}(t)\right]d \lambda}{\sigma T_{\rm bb}(t)^4} \, ,
\end{equation}

where $B_{\rm \lambda}$ 
%$= \left(2 \pi h c^2/\lambda^5\right)   \left[{\rm exp}\left(\displaystyle \frac{hc}{\lambda k_B T_{bb}}\right)-1\right]^{-1}$ i
is the blackbody spectral radiance per~unit wavelenght, $\sigma $ the Stefan-Boltzmann constant and~$T_{\rm bb}$ the blackbody temperature. The latter is related to the bolometric luminosity by  $L(t)= 4 \pi \sigma R_{\rm ph}(t)^2 T_{\rm bb}^4(t)$~, where the photospheric radius $R_{\rm ph}(t)=R(t)-2V(t)/(3kM)$.

Dust extinction is very pronounced at UV wavelengths \citep{cardelli}, being characterized by a bump at 220 nm, close to the ULTRASAT band.
We adopt the averaged dust extinction model of \cite{cardelli} for diffuse interstellar medium in the Milky Way with RV = 3.1 from $\lambda_{\rm min} = 0.1 ~\mu$m to $\lambda_{\rm max} =~3.3~\mu$m.~We account for the variation of total extinction with wavelength, $\eta_{\rm ext}$ using $A_{\lambda}=1.086~\tau_{ext}$, and compute the resulting attenuation in the ULTRASAT band, 

\begin{equation}
  \eta_{\rm ext, \rm UV}= \frac{\int_{\lambda_{min}}^{\lambda_{max}}e^{-\tau_{ext}}}{\Delta \lambda}\sim 0.13  \, .
\end{equation}

\begin{figure}
    \centering
    \includegraphics[width=1.1\linewidth]{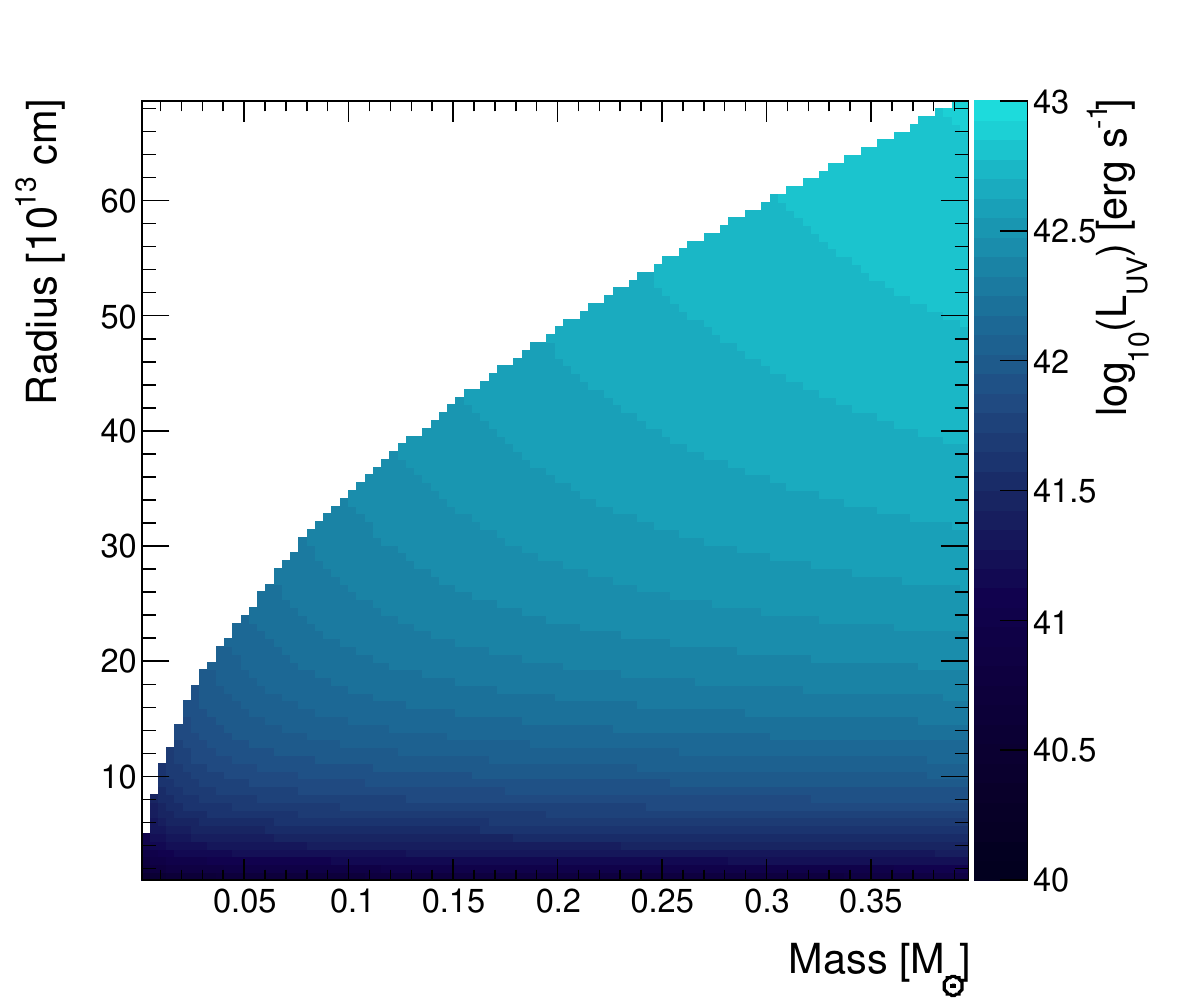}
    \caption{UV-band luminosity of sources with CSM masses from $10^{-3}\,M_\odot$ to $0.4\,M_\odot$ and inner shell radii between $10^{13}$~cm and $10^{15}$~cm, including the effects of Galactic extinction.}

    \label{lumi_UV}

    \includegraphics[width=1.1\linewidth]{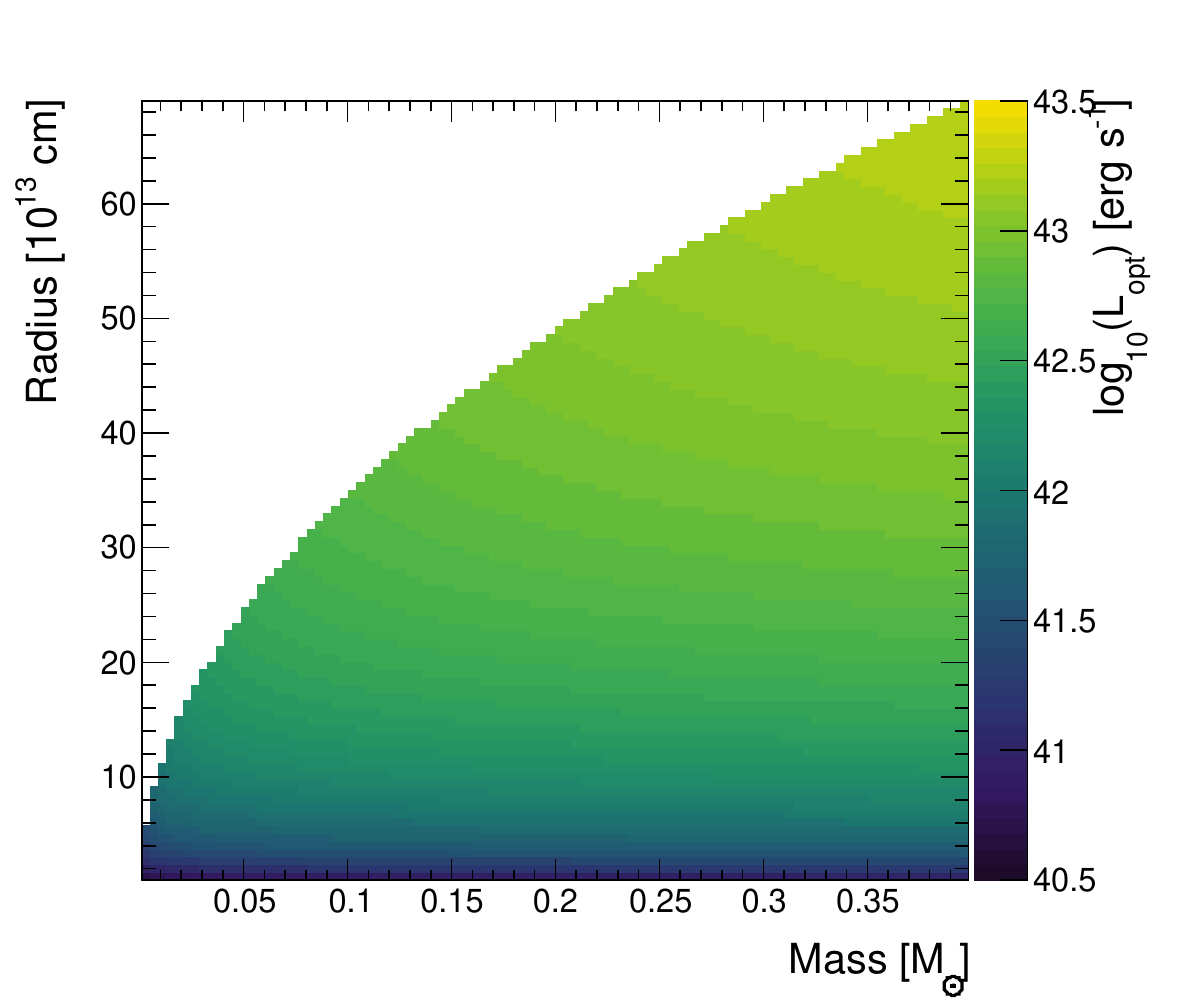}

    \caption{Optical-band luminosity of sources with CSM masses from $10^{-3}\,M_\odot$ to $0.4\,M_\odot$ and inner shell radii between $10^{13}$~cm and $10^{15}$~cm, including the effects of Galactic extinction.}

    \label{lumi_opt}
\end{figure}

The UV-band luminosity as a function of the CSM mass and radial extent, accounting for extinction, is shown in Figure~\ref{lumi_UV}. After correcting for extinction of the emission, i.e. after multiplying $L_{UV}(t)$ by eq. 3, we assessed the detectability of our model SN-CSM interaction events by comparing the resulting flux, as a function of redshift, to the sensitivity of either ULTRASAT and  ZTF, expressed as their respective limiting magnitudes in the AB system, $m_{\rm AB}^{\rm lim}$. The limiting magnitude is related to the minimum detectable spectral flux density through the relation:

\begin{equation}
    \label{eq:mab_fnu}
    m^{\rm lim}_{\mathrm{AB}}=-2.5\log_{10}\left( \frac{f_{\nu}}{\rm [Jy]}\right)+8.90,
\end{equation}
that can in turn be converted into the corresponding spectral flux density per unit wavelength $f_{\lambda}$ as:
\begin{equation}
\label{eq:fnu_flambda_conversion}
    \frac{f_{\nu}}{\rm [Jy]}=3.34\times 10^4 \left(\frac{\lambda}{\rm \AA}\right)^2 \frac{f_{\lambda}}{\rm [erg~cm^{-2}~s^{-1}~\AA^{-1}]},
\end{equation}
where $\lambda$ is the so-called \textit{pivot} wavelength, i.e. a measure of the effective wavelength of a filter.~In the case of ULTRASAT we consider the average wavelength in its observing band ($\bar{\lambda}=260$~nm), while for ZTF we directly use values from the SVO Filter Profile Service\footnote{\url{http://svo2.cab.inta-csic.es/svo/theory/fps/}} \citep{svo1,svo2}. For both instruments, we obtain $f_{\lambda}^{\rm lim}\simeq 1.6\times 10^{-9}~\mathrm{erg~cm^{-3}~s^{-1}}$. 

We define the average instantaneous flux of the source as a function of redshift, CSM shell mass, and radius as:

\begin{equation}
    f(z,M,R)= \frac{1}{\Delta t} \int _0 ^{\Delta t} \frac{f_{UV}(t,z) L(t, M, R)e^{-\tau_{ext}} }{4\pi d_L(z)^2 \Delta \lambda } dt 
    \label{eq:f}
\end{equation}

We consider a distribution for the CSM radius (R) and mass (M) based on typical stellar population models, with a focus on a specified range of values for both parameters as reported, e.g., in \cite{margalit_2022}. We compare this quantity (eq. \ref{eq:f}) to the minimum detectable instantaneous flux of ULTRASAT, $f_{\lambda}^{\rm lim}$. For a representative pointing, if the source maintains an instantaneous flux above this threshold during the exposure time $\Delta t=300$~s \citep{ultrasat}, it will be detectable once integrated over that time. 

\begin{figure}
    \centering
     \includegraphics[width=1.1\linewidth]{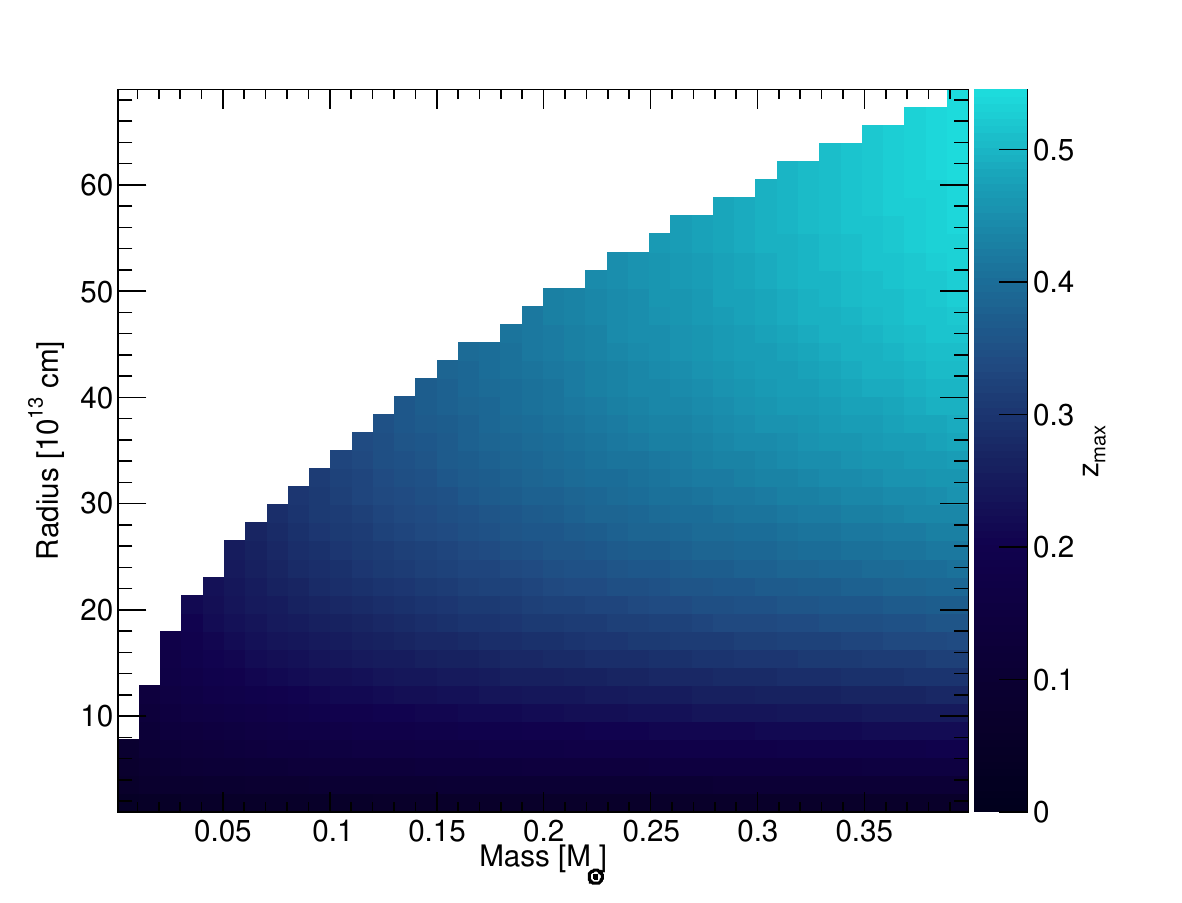}
     \caption{Maximum redshift detectable by ULTRASAT as a function of the inner CSM shell radius ($R$) and CSM mass ($M$). The color scale (z-axis) indicates the corresponding redshift. The radius ranges from $10^{13}$ to $10^{15}$~cm, while the CSM mass varies between $10^{-3}$ and $0.4\,M_{\odot}$. Absorption effects have been included in the calculation.}

    \label{fig:final}
\end{figure}

Figure \ref{fig:final} shows the dependence of the maximum observable redshift of a given source on the mass and radius.~Our results indicate that SBOs from CCSNe in the CSM scenario, can be detected by ULTRASAT out to a maximum distance $D_L \sim  3194 $ Mpc (z=0.55).

We applied the same procedure to the ZTF observing band ($\lambda_{\rm min}$ = 370 nm; $\lambda_{\rm max}$= 895 nm), estimating the SBO emission, from the same model events, which could be detected by the ZTF.~Besides the change in observing band, hence  $f_{UV}(t)$ becoming $f_{ZTF}(t)$, extinction is less pronounced in the optical bands and we estimated $ \eta_{\rm ext, \rm ZTF}= \sim 0.46$. The same analysis performed for the UV band has been carried out for the optical band as well. Figure~\ref{lumi_opt} shows the optical-band luminosity as a function of the CSM mass and radius, including the effects of extinction.

\begin{figure}
    \centering
    \includegraphics[width=1.1\linewidth]{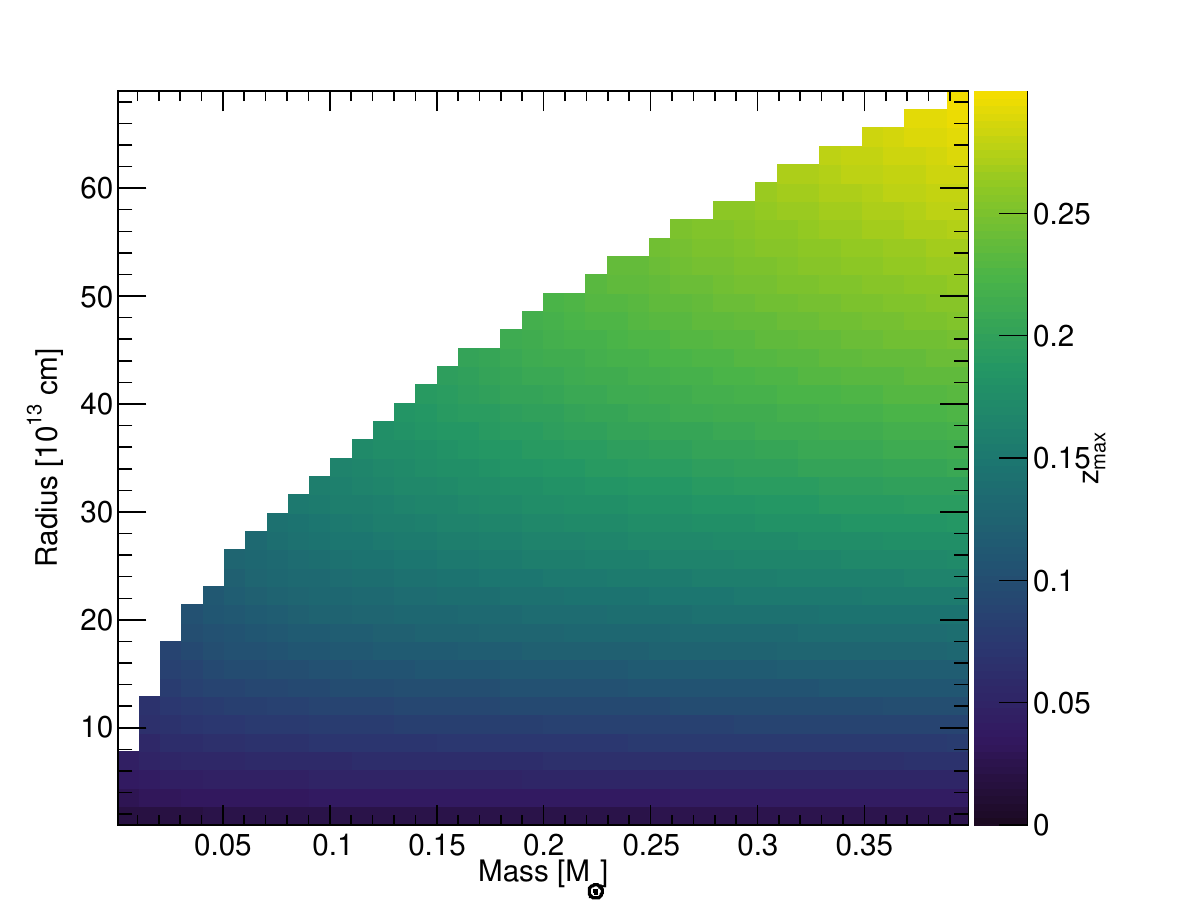}
    \caption{Maximum redshift detectable by ZTF as a function of the inner CSM shell radius ($R$) and CSM mass ($M$). The color scale (z-axis) indicates the corresponding redshift. The radius ranges from $10^{13}$ to $10^{15}$~cm, while the CSM mass spans from $10^{-3}\,M_{\odot}$ to $0.4\,M_{\odot}$. Absorption effects have been included in the calculation.}

    \label{fig:ztf2}
\end{figure}

The figure \ref{fig:ztf2} shows the horizon distance as function of the mass and radius from observations made by ZTF.
Our results indicate that SBOs from CCSNe in the CSM scenario, can be detected by ZTF out to a maximum distance $D_L \sim 1562  $ Mpc (z=0.30).

\section{Expected detection rate}

\label{sec:rate}

In order to estimate an expected rate of detections we must average the function $z_{\rm lim} (M, R)$  over the  probability distributions of $R$- and $M$-values within the population. Due to the limited observational data for CCSNe surrounded by CSM, the actual distributions are not known, although the observed sources point to relatively wide ranges both in shell masses, $\sim (10^{-3} \div 0.4) M_\odot$, and radii, $\sim (10^{13} \div 10^{15})$ cm \citep{margalit_2021}.~We adopted these as fiducial ranges while choosing an agnostic approach to model the probability distribution of $M$ and $R$ within such limits.~In particular, we tried three different assumptions:(i) a uniform probability distribution, (ii) a scale-invariant (log-flat) distribution in logarithmic space, defined as 
$P(M) = k_M / M$ and $P(R) = k_R / R$, and (iii) a normal distribution. 

For each of the three assumptions, we integrate the specific rate of core-collapse events up to $z_{\rm lim} (M, R)$ for ULTRASAT, and then marginalize over the corresponding probability distributions of $M$ and $R$, in order to estimate a population-averaged  rate of detections (per year).~In formulae: 

\begin{multline}
    \dot{N}= \int_{M_{\rm min}}^{M_{\rm max}} dM 
    \int_{R_{\rm min}}^{R_{\rm max}} dR ~\times \\  
    \int_0^{z_{\rm lim}(M,R)} \frac{R_{CCSN}(z)}{1+z} \frac{dV(z)}{dz} P(M) P(R) dz
     \label{eq:rate1}   
\end{multline}

where $R_{CCSN}(z)=k_{CC} \times \psi(z)$ is the comoving volumetric rate of SNe, $k_{CC}=0.0068~M_{\odot}^{-1}$ being the number of stars that explode as SNe per unit mass given a Salpeter initial mass function (IMF, \citealt{Salpeter_1955}), and 
$\psi(z)= 0.015 ~\displaystyle \frac{(1+z)^{2.7}}{1+ \left[\left(1+z\right)/2.9\right]^{5.6}}~M_{\odot} ~{\rm yr}^{-1} ~{\rm Mpc}^{-3}$ describes the cosmic star formation history. ~The comoving volume $dV(z)=D^3_{H} d^2_C(z)/F(z)~dz d \Omega$, where $D_H = c/H_0$ is Hubble distance, the proper distance 
\begin{equation}
    d_C(z) = \int_0^z \frac{1}{F(z')}dz' \, ,
\end{equation}
and we defined
\begin{equation}
    F(z)= \sqrt{\Omega_M (1+z)^3 + \Omega_k(1+z)^2 + \Omega_{\Lambda}} \, .
\end{equation}

The cosmological parameters used are $\Omega_M= 0.685$, $\Omega_{\Lambda} = 0.315$, and $\Omega_k=0$ \citep{Aghanim_2020}.

The Zwicky Transient Facility (ZTF) has an instantaneous field of view of approximately 47~deg$^2$ and features the largest cameras among telescopes with apertures greater than 0.5 meters.~The limiting redshift function \( z_{\rm lim}(M, R) \) reaches a maximum value of \( z_{\rm max} = 0.30 \) within the parameter space explored. This defines the furthest distance at which any of the sources considered here would be detectable by ZTF.

~Although in this work we did not attempt a systematic population study, we find that, when computing the expected detection rate, assumptions (i) and (ii) described above hardly produce results that are consistent with the population observed by ZTF.~On the other hand, a normal distribution for both $M$ and $R$  provided a better agreement of predictions with ZTF observations, if we set the peaks of the two distributions respectively at \( 9 \times 10^{-2}\,M_{\odot} \) and \( 3 \times 10^{14}\,\mathrm{cm} \),  
and with standard deviation \( \sigma_M = 7 \times 10^{-3}\,M_{\odot} \) and \( \sigma_R = 5 \times 10^{13}\,\mathrm{cm} \).~This choice, which is clearly not unique, yields an event rate of $\sim 28 ~\mathrm{yr}^{-1}$, distributed as follows: about 25 events originate from progenitors with masses in the range $(10^{-3}\div 10^{-1}) M_{\odot}$, and 3 events originate from progenitors with \( (0.1\text{--}0.4)\,M_{\odot} \), which falls within the mass range predicted for FBOTs by \citet{Liu_2022},with a median ejecta mass  of \( M_{\rm ej} = 0.11^{+0.22}_{-0.09}\,M_{\odot} \).
ZTF has observed a total of 277 Type IIn supernovae to date, with 160 events passing the sample selection criteria \citep{Pessi_2025}. The resulting SN IIn detection rate by ZTF is estimated to be $R_{\rm ZTF, obs} \sim 23~\mathrm{yr}^{-1}$, which is of the same order as our estimated rate. Note that this observed rate corresponds to about 12\% of the total core-collapse supernovae (CCSNe) rate expected within the maximum volume sampled by ZTF, i.e., up to \( z_{\max} \sim 0.30 \), as discussed above.

We adopted the same normal priors to estimate the expected ULTRASAT detection rate for CSM-interacting type-IIn SNe. Considering ULTRASAT’s high-cadence mode, in which the telescope observes for approximately 21 hours per day, we obtain an expected detection rate

We adopted the same normal priors to estimate the expected ULTRASAT detection rate for CSM-interacting type-IIn SNe. Considering ULTRASAT’s high-cadence mode, in which the telescope observes for approximately 21 hours per day, we obtain an expected detection rate $R_{\rm CCSNe, \rm ULTRASAT} \sim  3416 ~\mathrm{yr}^{-1}$.~For the FoV, we assume that ULTRASAT will operate a UV telescope with an unprecedented \(204~\mathrm{deg}^2\). The simulated events are distributed as follows: 3085 events have ejecta masses in the range \(10^{-3} - 0.1~M_{\odot}\), while 331 events fall within the range \(0.1 - 0.4~M_{\odot}\).

\section{High energy neutrino signatures}

\label{sec:neutrino}
As outlined in the previous section, CSM-interacting SNe are %regarded as 
potential neutrino emitters.~The neutrino luminosity produced by shocks propagating into a typical stellar wind or interstellar medium (ISM) can last for several hundred days, however, the resulting flux is generally too weak to be detected by Icecube and KM3NeT. %significantly to the diffuse neutrino background.
~This emission is accompanied by a comparable \(\gamma\)-ray luminosity, as the emission originates at large radii (\(> 10^{16}\) cm) \citep{Waxman_2024} and, thus, the pair production optical depth remains low for most of its duration.~A dense and optically thick CSM surrounding the progenitor, on the other hand, can give rise to a more intense neutrino emission.~This occurs around the time of the supernova SBO, as the shock transitions from a radiation-mediated to a collisionless regime while traversing the CSM shell (see also \citealt{Li_2019}).

High-energy protons, accelerated at the SN shock front, can interact inelastically with cold nucleons in the CSM. The main  role is played by p-p collisions, leading to the production of mesons, predominantly charged and neutral pions. While neutral pions $\pi^0$ decay into $\gamma$-rays, charged pions $\pi^+/ \pi^-$ decay into muons and subsequently into high-energy neutrinos (see  \citealt{Murase_2011}, which explores high-energy neutrino emission from extragalactic cosmic-ray sources, including scenarios relevant to supernovae IIn.)\\
In order to estimate the expected neutrino detection rate, we start by calculating the emission spectrum. Following \cite{Waxman_2024}, for an accelerated proton spectrum described by
\[
\epsilon_p^2 \frac{dn_p}{d\epsilon_p} = \mathrm{const},
\]
over six decades in energy, from 1~GeV to 10$^4$~TeV,~the resulting neutrino  spectrum is flat in $E_\nu^2~ dn_\nu/dE_\nu$, and can be approximated by:
\begin{equation}
    E_{\nu}^2 \frac{dn_{\nu}}{dE_{\nu}} \sim 0.5 \epsilon_{\rm CR} L_{\rm br}t_{\rm br}, 
\end{equation}
where $L_{\rm br}$ is the breakout luminosity at $t=0$ and we assume %\(L_{\rm br} = L(0)\) with 
\(t_{\rm br} = t_0\). The parameter \(\epsilon_{\rm CR}\) denotes the fraction of the shock breakout luminosity \(L_{\rm br}\) converted into accelerated protons.
The neutrino fluence \(\Phi(E_{\nu})\) is defined as the time-integrated neutrino energy flux at Earth, and can be written as
\begin{equation}
    \Phi(E_{\nu}) = \frac{1}{4 \pi d^2} \int_0^{t_{0}} L_{\nu}(E_{\nu}, t) \, dt,
    \label{eq:fluence}
\end{equation}
where \(L_{\nu}(E_{\nu}, t)\) is the neutrino spectral luminosity of the source at energy \(E_{\nu}\) and time \(t\), \(d\) is the distance to the source, and \(t_{\rm SBO}\) is the characteristic timescale of the SBO emission.

Using this fluence, the expected number of neutrino events \(N_{\nu}\) in a detector with energy-dependent effective area \(A_{\rm eff}(E_{\nu})\) is given by

\begin{equation}
    N_{\nu} = \int_{E_1}^{E_2} \frac{\Phi(E_{\nu})}{E_{\nu}^2} A_{\rm eff}(E_{\nu}) \, dE_{\nu}.
    \label{eq:N_nu}
\end{equation}

Here, \(A_{\rm eff}(E_{\nu})\) represents the effective area for muon neutrino (\(\nu_{\mu}\)) track events as a function of energy, taken from IceCube \citep{art:icecube_aeff} and KM3NeT \citep{art:loi}. The integration limits are set to \(E_{1} = 1~\mathrm{TeV}\) and \(E_{2} = 300~\mathrm{TeV}\).

Given a CSM shell mass \( M \) and a radius \( R \), and the luminosity \( L_{\rm br} \) expressed as a function of these parameters, we calculate the expected neutrino luminosity  adopting a proton acceleration efficiency  \(\epsilon_{\mathrm{CR}} = 0.1\) \citep{Waxman_2024},  and then compute the expected number of neutrino events as a function of distance.~ The  threshold of at least 1 observed event determines the maximum detection redshift.
%\textbf{LA SOGLIA DI OSSERVAZIONE DI NEUTRINI ORA è 1 EVENTO}

\begin{figure}
    \centering
    \includegraphics[width=1.1\linewidth]{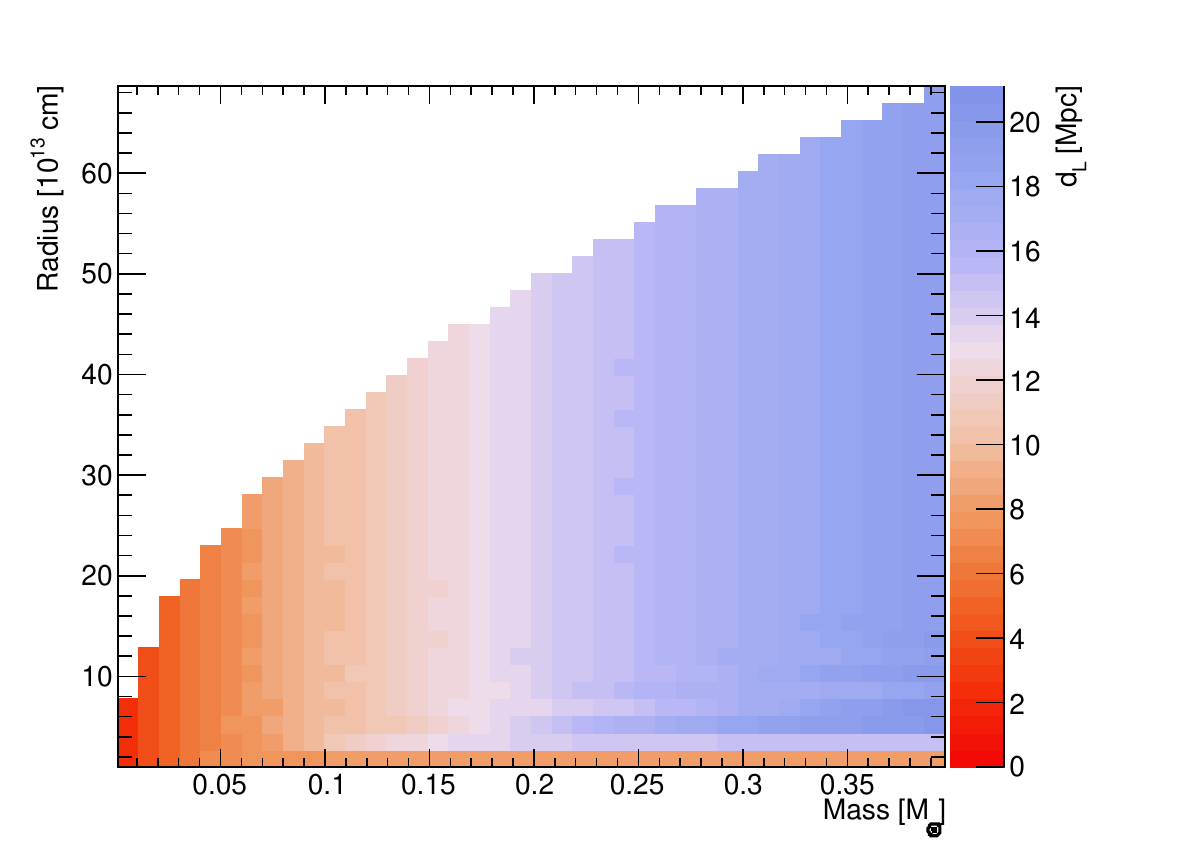}
    \caption{Maximum redshift detectable by IceCube as a function of the inner CSM shell radius ($R$) and CSM mass ($M$). The color scale (z-axis) indicates the corresponding redshift. The radius ranges from $10^{13}$ to $10^{15}$~cm, while the CSM mass varies between $10^{-3}\,M_{\odot}$ to $0.4\,M_{\odot}$.}

    \label{icecube}

\end{figure}

\begin{figure}
    \includegraphics[width=1.1\linewidth]{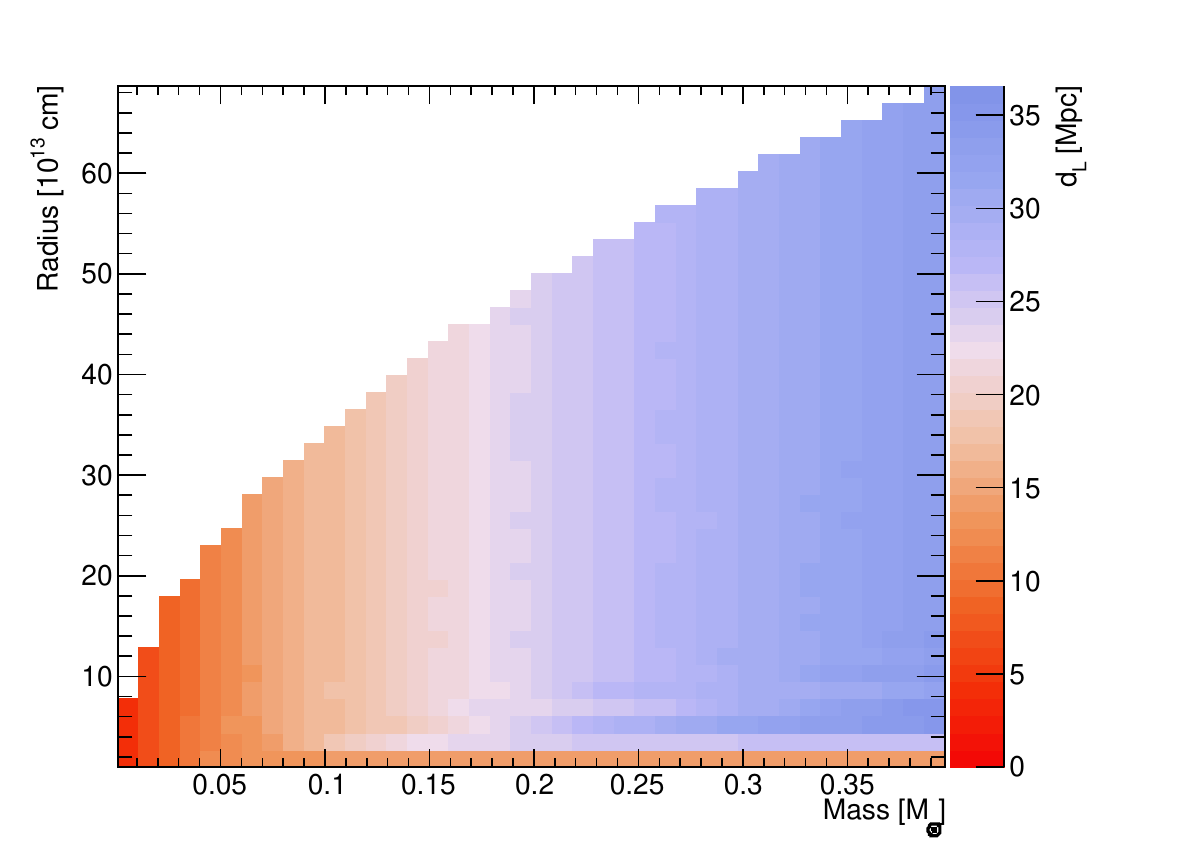}
    \caption{Maximum redshift detectable by KM3NeT as a function of the inner CSM shell radius ($R$) and CSM mass ($M$). The color scale (z-axis) indicates the corresponding redshift. The radius ranges from $10^{13}$ to $10^{15}$~cm, while the CSM mass varies between $10^{-3}\,M_{\odot}$ to $0.4\,M_{\odot}$.}

    \label{fig:km3}
\end{figure}

Our results, shown in Fig.~\ref{icecube} and \ref{fig:km3}, indicate that the maximum distance at which neutrinos emitted by CSM-interacting SNe can be detected is $ D_{L,{\rm KM3}} \sim 33.4~{\rm Mpc}$ for KM3Net, and $ D_{L,{\rm Ice}}\sim 19.5$~Mpc with  Icecube.~Using the same prior as in Sec.~\ref{sec:rate} to represent the cosmic source population, we obtain an IceCube event rate of \( R_{\mathrm{Ice}} = 0.35~\mathrm{yr}^{-1} \) and \( R_{\mathrm{KM3}} = 1.78~\mathrm{yr}^{-1} \) in the case of KM3NeT, suggesting promising chances for detection particularly with the latter instrument.~The CCSNe contributing to the expected detections all exhibit ejecta masses greater than 0.1~\(M_\odot\), resulting in UV/optical/bolometric luminosities consistent with FBOTs. This supports the idea that FBOTs may represent a sub-class of CCSNe and  promising sources of high-energy neutrinos \citep{Guarini_2022}.

Using the well-studied transient AT2018cow as a reference case, we perform a calculation to estimate the maximum distance at which such a source could be detected by ULTRASAT and by current neutrino observatories. The source, described in \citet{Prentice_2018}, is characterized by a high peak luminosity ($\sim 1.7 \times 10^{44}~\mathrm{erg~s^{-1}}$), a hot blackbody spectrum peaking at $\sim27,000~\mathrm{K}$, and an estimated ejecta mass of $0.1$–$0.4~M_\odot$. Based on these parameters, a transient similar to AT2018cow could be detectable by ULTRASAT up to a maximum distance of $\sim1.5~\mathrm{Gpc}$ ($z \approx 0.28$).~Moreover, our calculations indicate that the maximum distance at which at least one neutrino could be detected from a source with the peak luminosity of AT2018cow is %calculations show that this distance is approximately  
$\sim D_{\rm max,IceCube} \sim (9.8-19.4)~\mathrm{Mpc}$,  %and $D_{\rm max,IceCube}\sim 19.4~\mathrm{Mpc}$, respectively, 
assuming %the standard IceCube sensitivity and 
a CSM-shell mass $M= (0.1-0.4) M_{\odot}$.~Adopting the KM3NeT sensitivity, on the other hand, we find 
$D_{\rm max,KM3NeT}\sim 17.0~\mathrm{Mpc}$ for $M= 0.1 M_{\odot}$ and $D_{\rm max,KM3NeT}\sim33.7 ~\mathrm{Mpc}$ for $M= 0.4 M_{\odot}$.

%Fig.~\ref{fig:ATCOW} reports the expected number of neutrino events vs. \st{unit} energy for AT2018cow, that is at $D\sim 60 $ Mpc, considering the effective area of KM3NeT. In Fig.~\ref{fig:ATCOW_15Mpc}  we show the neutrino event number per unit energy for AT2018cow-like source located at   $D=D_{\rm max,KM3NeT}\sim 17 $ Mpc.

\begin{figure}
    \centering
    \includegraphics[width=1.1\linewidth]{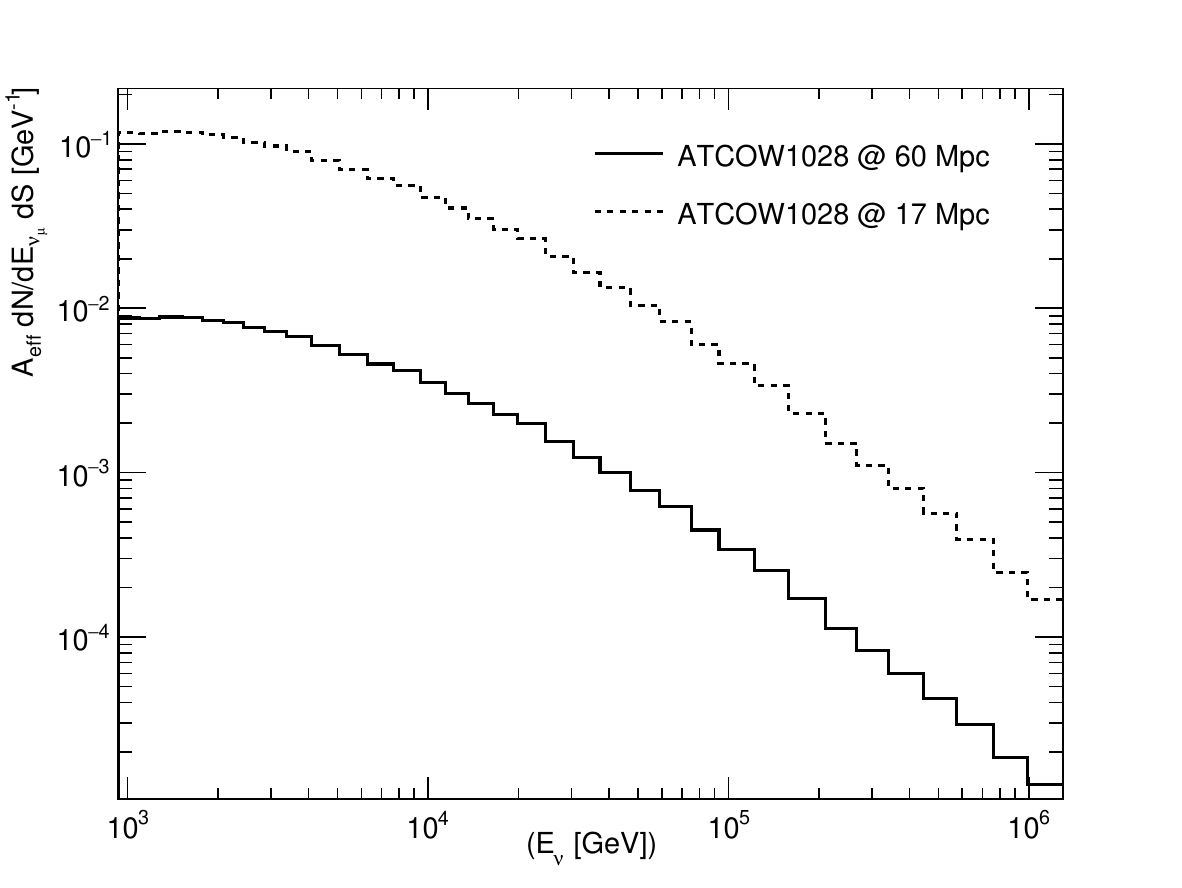}
    \caption{Neutrino event number per unit energy for KM3NeT from an AT2018cow-like source. 
    The solid curve corresponds to a source located at $D \sim 60$~Mpc, while the dashed curve represents the same source placed at 
    $D = D_{\rm max,KM3NeT} \sim 17$~Mpc, i.e. the maximum distance at which KM3NeT can detect a neutrino from this source.}

    \label{fig:ATCOW}
\end{figure}

Fig.~\ref{fig:ATCOW} shows the expected number of neutrino events per unit energy for an AT2018cow-like source, obtained using the effective area of KM3NeT. 
The solid curve corresponds to AT2018cow at its actual distance of $D \sim 60$~Mpc, while the dashed curve represents the same source placed at 
$D = D_{\rm max,KM3NeT} \sim 17$~Mpc, which is the maximum distance at which KM3NeT is expected to detect at least one neutrino from such an event.

\section{Discussion and conclusions}

\label{sec:discussion}

In CCSNe interacting with CSM, luminosity variations are primarily driven by the mass and spatial distribution of the CSM material. 

In this work, we calculate the bolometric and UV light curves associated to SBO from a CSM shell, characterized by its mass (\( M \)) and radius (\( R \)),  adopting the
analytical expressions by \cite{margalit_2022}.~Accounting for the $M$- and $R$-dependence of the predicted light curves, we estimate the maximum observable horizon as a function of these parameters, and then estimate the expected detection rate of such events by the ULTRASAT satellite, averaging over an agnostic distribution of ($M$,$R$)-values within the population.

Our results show that ULTRASAT will be capable of detecting such events out to   \( z \sim 0.55 \) (\( D_L \sim 3 \)~Gpc), whereas ZTF was limited to to \( z \sim 0.30 \) (\( D_L \sim 1.5 \)~Gpc). These findings can be directly compared to the results of \citep{art:zegarelli}, who found that the farthest distance at which a UV (optical) signal from the cooling emission following an SBO in red supergiants (RSGs) can be detected by ULTRASAT is \( z \sim 0.24 \) (about 1~Gpc), and by ZTF up to \( z \sim 0.19 \) (approximately 800~Mpc).
Our estimated event rate for ZTF, $\sim 28$ per year, compares well with their observed type~IIn SN rate $\sim$ 23~per year \citep{Pessi_2025}, which are characterized by strong spectroscopic evidence of CSM interaction.~For ULTRASAT, our expected detection rate is significantly higher, \(\sim 3500 \)~yr$^{-1}$, and results from a $\sim 5$ times wider FoV, 4 times larger duty cycle \citep{art:zegarelli} and improved sensitivity which roughly doubles the horizon reach.
Future observational studies of the relevant SN populations will help refine these predictions and improve our understanding of early CSM interaction signatures.

%As highlighted in previous work on choked jet transients \citep{art:zegarelli}, the temporal availability of different observatories must be taken into account. 
Although ZTF is unlikely to be operational at the time of ULTRASAT’s launch, it has been used here to calibrate and validate the expected detection rates for ULTRASAT. Therefore, the comparisons presented serve as a sensitivity benchmark rather than implying simultaneous multi-wavelength coverage.

%Since CSM-interacting SNe are considered to be potential neutrino emitters, it is useful, within a multi-messenger framework, to estimate the expected neutrino luminosity, the implied detection horizon and the corresponding neutrino detection rate. We find a maximum deetection distance \( D_{L,\mathrm{KM3}} \sim 18 \)~Mpc for KM3NeT, and \( D_{L,\mathrm{Ice}} \sim 9 \)~Mpc in the case of IceCube. Based on these values and on our population model, we obtain an estimated event rate of \( R_{\mathrm{Ice}} = 0.3 \)~yr\(^{-1} \) for IceCube and \( R_{\mathrm{KM3}} = 1.8 \)~yr\(^{-1} \) for KM3NeT. 

CSM-interacting SNe are considered potential neutrino emitters and early UV detections with ULTRASAT could serve as effective triggers for high-energy neutrino observatories. By providing rapid alerts within a narrow time window around the explosion, ULTRASAT observations would enable time-coincident searches for neutrinos, substantially improving the sensitivity of the analysis by reducing the background rate. These considerations prompted us to estimate the expected neutrino luminosity of CSM-interacting SNe, and their corresponding detection horizon and rates.~We find a maximum horizon of \(D_{L,\mathrm{KM3}} \sim 33~\mathrm{Mpc}\) for KM3NeT, and \(D_{L,\mathrm{Ice}} \sim 19~\mathrm{Mpc}\) in the case of IceCube. Based on these values and on our population model, the estimated event rate of \(R_{\mathrm{Ice}} = 0.35~\mathrm{yr}^{-1}\) for IceCube and \(R_{\mathrm{KM3}} = 1.78~\mathrm{yr}^{-1}\) for KM3NeT. 
Moreover, we analyzed the expected events per year as a function of CSM mass in our model calculations, finding that 25 events have a masses below 0.1~\( M_\odot \) while only 3 events for higher masses thus suggesting that low-mass CSM environments are more common.~Despite being rarer, high-mass CSM cases may nevertheless produce the more luminous transients, potentially consistent with the class of fast blue optical transients (FBOTs), and provide the dominant contribution to the   observed neutrino events.

Recent detections of the nearby SN~2023ixf and SN~2024ggi, both classified as Type IIn supernovae \citep{Jacobson-Galán_2023, Jacobson-Galán_2024}, have highlighted the prospects for future detections of similar events,  particularly with current wide-field surveys.~Previous studies \citep{Murase_2011, waxman17, Guetta_2023} have already explored the potential for neutrino emission from such sources which, despite their relative proximity, remain challenging to detect due to the limited neutrino horizon.

A promising direction for future investigation is a detailed analysis of the X-ray emission produced during the transition from a radiation-mediated to a collisionless shock, as discussed by \citet{Wasserman_2025}. This phase may carry crucial signatures of the shock physics and the structure of the CSM.

ULTRASAT-detected CCSNe will be matched with neutrino event catalogs to search for temporal and spatial coincidences. In this paper we  emphasize that early UV detection can serve as a predictive tool for neutrino observatories.

Alert Protocols: We propose integrating ULTRASAT detections into real-time alert systems to trigger targeted neutrino searches during the critical early post-explosion window.

%\textcolor{magenta}{Too neutrino e ULTRASAT da aggiungere}

%\textcolor{red}{Come fatto nel paper dei choked, commenterei che ZTF non sarà operativo quando ULTRASAT verrà lanciato e che quindi questi risultati vanno presi come riferimento ma probabilmente non ci sarà overlap tra i due ESATTO METTI ZTF-LIKE TELESCOPES COME HA FATTO ANGELA}

\section{Acknowledgement}
This research was partially funded by the Israel Ministry of Innovation, Science, and Technology through Grant number 0008107.

\clearpage

\bibliography{biblio.bib}{}
\bibliographystyle{aasjournal}

\end{document}